\documentclass[twocolumn,showpacs,preprintnumbers,amsmath,amssymb,epsfig,widetext]{revtex4}
\usepackage{graphicx}
\usepackage{dcolumn}
\usepackage{bm}
\usepackage{epsfig}
\usepackage{color}

\def\fun#1#2{\lower3.6pt\vbox{\baselineskip0pt\lineskip.9pt
  \ialign{$\mathsurround=0pt#1\hfil##\hfil$\crcr#2\crcr\sim\crcr}}}
\def\simgt{\mathrel{\lower0.6ex\hbox{$\buildrel {\textstyle >}
 \over {\scriptstyle \sim}$}}}
\def\simlt{\mathrel{\lower0.6ex\hbox{$\buildrel {\textstyle <}
 \over {\scriptstyle \sim}$}}}

\input epsf

\newcommand{\hompc}{\,h\,{\rm Mpc}^{-1}}
\newcommand{\mpcoh}{\,h^{-1}\,{\rm Mpc}}

\def\be{\begin{equation}}
\def\ee{\end{equation}}
\def\ba{\begin{eqnarray}}
\def\ea{\end{eqnarray}}

\def\nn{\nonumber}

\begin{document}

\preprint{}

\title{Chasing Unbiased Spectra of the Universe}
 
\author{Yong-Seon Song$^1$, Takahiro Nishimichi$^2$, Atsushi Taruya$^3$, Issha Kayo$^4$}
\email{ysong@kasi.re.kr; takahiro.nishimichi@ipmu.jp; ataruya@utap.phys.s.u-tokyo.ac.jp; kayo@ph.sci.toho-u.ac.jp}
\affiliation{
$^1$Korea Astronomy and Space Science Institute, Daejeon 305-348, R. Korea\\
$^2$Kavli Institute for the Physics and Mathematics of the Universe, Todai Institutes for Advanced Study, 
the University of Tokyo (Kavli IPMU, WPI), Kashiwa, Chiba 277-8583, Japan\\
$^3$Research Center for Early Universe, School of Science, University of Tokyo, Bunkyo-ku, Tokyo 113-0033, Japan\\
$^4$Department of Physics, Toho University, 2-2-1 Miyama, Funabashi, Chiba 274-8510, Japan}

\date{\today}

\begin{abstract}
The cosmological power spectrum of the coherent matter flow is
measured exploiting an improved prescription
for the 
apparent anisotropic clustering pattern 
in redshift space. New statistical
analysis is presented to provide an optimal observational platform to
link the improved redshift distortion theoretical model to future real
datasets. 
The statistical power as well as robustness of our method are 
tested against 60 realizations
of $8\,h^{-3}$ Gpc$^3$ dark matter simulation maps mocking the precision
level of upcoming wide--deep surveys. 
We showed that we can accurately extract the velocity power spectrum
up to quasi linear scales of $k\sim 0.1\hompc$ at $z = 0.35$
and up to $k\sim 0.15\hompc$ at higher redshifts within a couple of
percentage precision level. Our understanding of redshift space
distortion is proved to be appropriate for precision cosmology, and our
statistical method will guide us to righteous path to meet the real
world.
\end{abstract}

\pacs{draft}


\maketitle

\section{introduction}

The emergence of a standard model for the Universe dominated by an
unknown substance of dark materials has revolutionized our understanding
of the Universe. Since the first firm evidence of dark energy in 1998~\cite{Riess:1998cb,Perlmutter:1998np}, there has been substantial observational and theoretical research aiming at understanding the true nature of this phenomenon. In recent years, many authors have started exploring the possibility that dark energy, and the observed acceleration of the expansion of the Universe, could be the consequence of an incomplete theory of gravity on cosmological scales and may require modifications to Einstein's  theory of General Relativity. The information of underlying science about the Universe is given by looking at structure formation on large scales.

Several authors have shown that by combining various probes of the large--scale structure in the Universe, it is possible to test the relationship between these quantities which, in the linear regime, can generally be described by two functions of time and scale~\cite{Zhang:2007nk,Jain:2007yk,Song:2008vm,Reyes:2010tr,Song:2010fg,Yoo:2012vm}. Those can be constrained through a combination of weak lensing and an independent probe of matter--energy fluctuations. This is motivated by the fact that the weak lensing experiments probe the geometrical potential of combination between curvature perturbation and Newtonian force, which determines the trajectories of photons through the Universe, while matter fluctuation measurements probe the Newtonian force alone determining local inhomogeneities of matter--energy. 

The coherent motion of the galaxies opens a unique opportunity to access the fluctuations of the underlying 
matter density field~\cite{Guzzo:2008ac,Song:2008qt,Wang:2007ht,Percival:2008sh,White:2008jy,McDonald:2008sh,Jeong:2006xd,Jeong:2008rj,Desjacques:2009kt,Jennings:2010uv,Reid:2011ar,Okumura:2011pb,Kwan:2011hr,Samushia:2011cs,Blake:2011rj,Zhang:2012yt}.
This technique relies on the redshift space distortions seen in galaxy surveys. Even though we expect the clustering of galaxies in real space to have no preferred direction, galaxy maps produced by estimating distances from redshifts obtained in spectroscopic surveys reveal an anisotropic galaxy distribution. The anisotropies arise because galaxy recession velocities, from which distances are inferred, include components from both the Hubble flow and peculiar velocities driven by the clustering of matter~\cite{Kaiser:1987qv,Fisher:1994ks,Scoccimarro:2004tg}. Measurements of the anisotropies allow constraints to be placed on the rate of growth of clustering.
  
Measurements of coherent motion field from redshift distortion maps have been plagued by systematic uncertainties which have made their cosmological constraints uncompetitive compared to other probes of the Universe. The cosmological density and velocity field couple together and evolve nonlinearly.
In addition, the mapping formula between the real and redshift space is intrinsically nonlinear.
These nonlinearities prevent us from inferring the linear coherent motion from the redshift space clustering straightforwardly.
Recently, an accurate theoretical model for the redshift distortion was proposed in~\cite{Taruya:2010mx}.
They take into account the fact that
the linear squeezing and non--linear smearing effects on distorted maps are not separable to each other
and develop a more elaborate description than simple factorized formulation.
The derived correction terms at leading higher orders assist us to achieve better fit to simulated data. 
They also include non--linear corrections formulated using the closure approximation to predict the non--linear growth in density--density, density--velocity and velocity--velocity spectra. 
In this work, we assume a perfect cross--correlation between density and velocity fields at linear level to decompose 
the coherent motion spectra.
Thus we have two spectra to be decomposed; the linear density power spectrum and its velocity counterpart.

Despite all theoretical efforts, the exact form of FoG effect is unknown. We adopt most common functional form of FoG effect, such as Gaussian or Lorentzian, and parameterize FoG effect using one-dimensional velocity dispersion representing
the randomness of the motion which erases the correlation structure on small scales. The parameter space is extended to include this uncertainty of FoG effect in addition to the linear spectra of density and velocity fields, which were originally proposed by~\cite{Song:2010bk,Tang:2011qj}. We run MCMC routine to find best set of spectra and FoG parameter, and we find that coherent motion spectra are measurable at linear regime with good precision. Coherent motion spectra are measured at a couple of percentage accuracy with appropriate $k$ cut--off.

The paper is organized as follows. In Sec.~\ref{sec:models}, we begin by introducing suggested theoretical models of redshift distortions. We then describe the statistical method to extract the coherent motion spectra. Using this method, Sec.~\ref{sec:measure} presents our main results on the measurements of linear density and velocity spectra. In Sec.~\ref{sec:discussion}, the impact of wrong prior assumption on the decomposition of power spectra is discussed. Finally, we conclude in Sec.~\ref{sec:conclusion}. 

\section{Methodology}\label{sec:models}

We first highlight analytical models for the
redshift-space power spectrum. We then show our methodology of
reconstructing the linear density and velocity power spectra from the
two-dimensional power spectrum observed in redshift space.

\subsection{Analytical models for the power spectrum in redshift space}
\label{sec:modelsA}

At large scale, we expect that the density field as well as the velocity
field are small perturbations to the homogeneous universe. When the
higher-order contributions are negligibly small, the two-dimensional
power spectrum in redshift space, $\tilde{P}(k,\mu)$, can be expressed
as
\begin{equation}
\tilde{P}(k,\mu) = P_{\delta\delta}^{\rm lin}(k) + 2\mu^2 P_{\delta\Theta}^{\rm lin}(k) + \mu^4 P_{\Theta\Theta}^{\rm lin}(k),
\label{eq:Kaiser}
\end{equation}
where we denote by $\delta$ and $\Theta$ the density contrast and the
velocity divergence, with the latter defined by $\Theta \equiv
-(1+z)\nabla{\bf v}/H$.  The auto- and cross-power spectra of the two
fields in linear theory are expressed as $P_{ij}^{\rm lin}(k)$, with $i$
and $j$ being either $\delta$ or $\Theta$.  This formula describes the
effect of coherent velocity flow towards overdensity at large scale
(Kaiser effect).  Because of this effect, the clustering pattern in
redshift space is enhanced along the line-of-sight direction.  If the
gravitational law follows the general relativity, the three spectra are
not independent, but related as
\begin{eqnarray}
P_{\delta\Theta}^{\rm lin}(k) &=& f(z)P_{\delta\delta}^{\rm lin}(k),\\
P_{\Theta\Theta}^{\rm lin}(k) &=& f^2(z)P_{\delta\delta}^{\rm lin}(k),
\end{eqnarray}
where $f(z)={\rm d} \ln D_+(z) / {\rm d} \ln a$ is the growth rate parameter with $D_+$ being the linear growth rate. 
In this case, all the velocity information is contained in the parameter, $f(z)$, which is constant over wavenumber $k$.

However, the planned/ongoing galaxy redshift surveys as well as the existing large surveys mainly target weakly
nonlinear scale, where the feature of BAOs is prominent. Moreover, by appropriately modeling this regime,
we can in principle enlarge the range of wavenumber to be taken into account in the analysis, and improve the
constraints on the gravitational law. Thus, we have to somehow incorporate nonlinearity to
make maximum use of these surveys. 

First of all, the cross- and auto-power spectra of the density and the velocity fields
are naturally expected to receive nonlinear corrections. 
Another important effect arises from the random motion of galaxies.
By combining these two effects, Scoccimarro~\cite{Scoccimarro:2004tg} proposes the following formula
\ba
\tilde{P}(k,\mu) = \left\{P_{\delta\delta}(k) + 2\mu^2 P_{\delta\Theta}(k) + \mu^4 P_{\Theta\Theta}(k)
\right\}G^{\rm X}(k\mu),
\label{eq:S04}
\ea
where the effect of random motion is captured by the factor $G^{\rm X}(k\mu)$.
Note that we have replaced the three linear spectra in Eq.~(\ref{eq:Kaiser}), $P_{ij}^{\rm lin}(k)$, with their 
nonlinear counterparts, $P_{ij}(k)$.
In the original paper by Scoccimarro~\cite{Scoccimarro:2004tg},
the factor $G^{\rm X}(k)$ is designed so that it accounts for the random motion of the galaxies at large scale,
and he considers the Gaussian shape for this factor:
\ba
G^{\rm X}(k\mu) = G^{\rm GAU}(k\mu) = \exp\left\{-(k\mu\sigma_{\rm v})^2\right\},
\label{eq:Ggauss}
\ea
where $\sigma_{\rm v}$ denotes the dispersion of the one-point PDF of the velocity in one-dimension.
At smaller scale, inside the cluster of galaxies, the virial motion of galaxies also give a suppression of
the power spectrum in redshift space.
This effect is called Finger-of-God (FoG), and can also be approximately described by multiplication of 
the factor $G^{\rm X}(k\mu)$.
A Lorentzian form of this factor has been frequently adopted based on the results of $N$-body simulations: 
\ba
G^{\rm X}(k\mu) = G^{\rm LOR}(k\mu) = \frac{1}{1+(k\mu\sigma_{\rm v})^2}.
\label{eq:Glorenz}
\ea
For convenience, we do not distinguish between these damping effects, and simply call them as FoG in this paper, 
although, strictly speaking, the former one at large scale has a different origin.

Recently, Taruya, Nishimichi \& Saito~\cite{Taruya:2010mx} proposed a more accurate model for the redshift-space distortion.
Motivated by the fact that the Kaiser and the FoG effects can not be separated, and should not be described as
a factorisable form as in Eq.~(\ref{eq:S04}), they derived correction terms which have long been overlooked.
Their formula reads
\ba
\tilde{P}(k,\mu) = \left\{P_{\delta\delta}(k) + 2\mu^2 P_{\delta\Theta}(k) + \mu^4 P_{\Theta\Theta}(k)\right.\nonumber\\
\left.+ A(k,\mu) + B(k,\mu)\right\}G^{\rm X}(k\mu),
\label{eq:TNS10}
\ea
where the full expressions for the terms $A(k,\mu)$ and $B(k,\mu)$ based on perturbation theory 
can be found in the Appendix A of that paper~\cite{Taruya:2010mx}.
Let us note some important features in the new terms. First, they include higher-order polynomials in $\mu$ 
and $f$: the $A$-term has a term which scales as $f^3\mu^6$, while we have a $f^4\mu^8$ contribution in the $B$-term.
Thus, they become relatively important at $\mu\simeq1$. Another point is that as these terms arise as a non-linear
coupling between the density and the velocity fields, they are of the order $\mathcal{O}(\{P_{ij}^{\rm lin}(k)\}^2)$.
We thus may able to omit them on linear scales.

\subsection{Decomposition strategy}
\label{sec:modelsB}

We now describe how we decompose the power spectrum in redshift space into spectra of density and velocity.
Before that, let us introduce two useful quantities controlling the amplitude of the linear power spectra.
We define
\ba
P_{\delta\delta}^{\rm lin}(k,z) = g_\delta^2(k,z)P_{\delta\delta}^{\rm lin}(k, z_{\rm lss}),\\
P_{\Theta\Theta}^{\rm lin}(k,z) = g_\Theta^2(k,z)P_{\Theta\Theta}^{\rm lin}(k, z_{\rm lss}),
\ea
where $z_{\rm lss}$ stands for the redshift at the time of the last scattering.
The parameters, $g_{\delta}$ and $g_{\Theta}$, describe the growth rate of the density and the velocity fields
from that epoch. Since the spectrum $P_{\delta\delta}^{\rm lin}(k,
z_{\rm lss})$ is well constrained by observations of the CMB
temperature anisotropy, the parameters solely capture the properties of the gravitational law, and are expected to be 
free from the assumptions in the initial condition.

In reconstructing the spectra, we first bin the power spectrum measured from simulations into bins of 
$k$ and $\mu$. Then, for the $i$-th bin of wavenumber $k$, which we denote $k_i$,
we estimate $P_{\delta\delta}^{\rm lin}(k_i)$, $P_{\delta\Theta}^{\rm lin}(k_i)$ and $P_{\Theta\Theta}^{\rm lin}(k_i)$ 
based on the $\mu$ dependence.
We assume that linear $\delta$ and linear $\Theta$ are perfectly correlated, 
$P^{\rm lin}_{\delta\Theta}=\sqrt{P^{\rm lin}_{\delta\delta}P^{\rm lin}_{\Theta\Theta}}$, and treat two parameters, 
$g_\delta(k_i)$ and $g_\Theta(k_i)$, as free parameters for the $k_i$-bin.
In fitting $\tilde{P}(k_i,\mu)$, we scale the spectra according to the parameters $g_\delta$ and $g_\Theta$:
\ba
P_{ij}^{\rm lin}(k) = \frac{g_i(k)}{g_i^{\rm fid}(k)}\frac{g_j(k)}{g_j^{\rm fid}(k)}P_{ij}^{\rm fid, lin}(k).\label{eq:p_scale}
\ea
where quantities with subscript ``fid'' are computed for the fiducial cosmological model.

When we restrict the analysis to linear regime, the above procedure is expected to work properly.
We further elaborate the procedure to correctly handle the nonlinearity.
We adopt the closure approximation (Taruya \& Hiramatsu~\cite{Taruya:2007xy}) to predict the nonlinear growth in the three spectra, 
$P_{ij}(k)$.
In doing so, we simply assume the fiducial GR cosmology used in running the simulations.
We pre-compute the nonlinear corrections to the three spectra up to the second-order in the Born approximation:
\ba\label{eq:closure}
\delta P_{ij}^{\rm fid}(k) = P_{ij}^{\rm fid}(k) - P_{ij}^{\rm fid, lin}(k),
\ea
while we allow to vary the linear part according to Eq.~(\ref{eq:p_scale}).
By adding up linear and nonlinear parts, we have
\ba
P_{ij}(k) = \frac{g_i(k)}{g_i^{\rm fid}(k)}\frac{g_j(k)}{g_j^{\rm fid}(k)}P_{ij}^{\rm fid, lin}(k)+\delta P_{ij}^{\rm fid}(k).
\ea
Strictly speaking, the second term depends on the cosmological model as well as the gravitational law.
As a first trial, however, we simply let this term unchanged from its fiducial value. 
In Sec.~\ref{sec:discussion}, we will relax the assumption to execute a more general analysis: we adopt a {\it wrong} cosmological 
model as the fiducial model, and see how well we can recover the {\it true} spectra.

We also scale the correction terms, $A(k,\mu)$ and $B(k,\mu)$ as follows.
In every step of the fitting, given set of $g_\delta(k)$ and $g_\Theta(k)$, we compute their simple arithmetic means:
\ba
\bar{g}_\delta = \frac{1}{N_{\rm bin}}\sum_i g_\delta(k_i),\qquad
\bar{g}_\Theta = \frac{1}{N_{\rm bin}}\sum_i g_\Theta(k_i),
\ea
where the subscript $i$ runs over $k$-bins up to a maximum wavenumber, $k_{\rm max}$, and we denote the number
of the bins by $N_{\rm bin}$. 
Again, we pre-compute the correction terms for the fiducial cosmology using the standard perturbation theory, 
which we denote $A^{\rm fid}$ and $B^{\rm fid}$, and scale them according to the average values of 
$g_\delta$ and $g_\theta$:
\ba\label{eq:AB}
A(k,\mu) &=& A^{\rm fid}\left(k,\mu;\frac{\bar{g}_\delta}{\bar{g}^{\rm fid}_\delta},
\frac{\bar{g}_\theta}{\bar{g}^{\rm fid}_\theta}\right),\nn\\
B(k,\mu) &=& B^{\rm fid}\left(k,\mu;\frac{\bar{g}_\delta}{\bar{g}^{\rm fid}_\delta},
\frac{\bar{g}_\theta}{\bar{g}^{\rm fid}_\theta}\right).
\ea
In the above, the terms originated from $\delta$ ($\Theta$) are multiplied by $\bar{g}_\delta / \bar{g}^{\rm fid}_\delta$ 
($\bar{g}_\Theta / \bar{g}^{\rm fid}_\Theta$). 

We finally explain our strategy for the FoG factor. We try both Gaussian and Lorentzian functions,
and we let $\sigma_v$ as a free parameter.
This parameter is determined by fitting globally the broadband shape of $\tilde{P}(k,\mu)$.

In summary, our reconstruction strategy is as follows.
We model the redshift-space power spectrum, $\tilde{P}(k,\mu)$, as
\ba\label{eq:Pth}
\tilde{P}(k,\mu) &=& \left\{\left[\left(\frac{g_\delta(k)}{g_\delta^{\rm fid}(k)}\right)^2
P_{\delta\delta}^{\rm fid, lin}(k)+\delta P_{\delta\delta}^{\rm fid}(k)\right]\right.
\nonumber\\
&&+\mu^2\left[\frac{g_\delta(k)}{g^{\rm fid}_\delta(k)}\frac{g_\Theta(k)}{g^{\rm fid}_\Theta(k)}
P_{\delta\Theta}^{\rm fid, lin}(k)+\delta P_{\delta\Theta}^{\rm fid}(k)\right]
\nonumber\\
&&+\mu^4\left[\left(\frac{g_\Theta(k)}{g_\Theta^{\rm fid}(k)}\right)^2P_{\Theta\Theta}^{\rm fid, lin}(k)+
\delta P_{\Theta\Theta}^{\rm fid}(k)\right]
\nonumber\\
&&\left.+A^{\rm fid}\left(k,\mu; \frac{\bar{g}_\delta}{\bar{g}^{\rm fid}_\delta}, \frac{\bar{g}_\Theta}{\bar{g}^{\rm fid}_\Theta}\right) 
+ B^{\rm fid}\left(k,\mu; \frac{\bar{g}_\delta}{\bar{g}^{\rm fid}_\delta}, \frac{\bar{g}_\Theta}{\bar{g}^{\rm fid}_\Theta}\right)\right\}
\nonumber\\
&&\times G^{X}(k\mu;\sigma_v),
\ea
where factors with the subscript ``fid'' are pre-computed for the fiducial cosmology, and $g_\delta(k_i)$ and
$g_\Theta(k_i)$ as well as $\sigma_v$ are free parameters.
We show results for various cases in what follows.
We include/exclude $A(k,\mu)$ and $B(k,\mu)$, we adopt Gaussian and Lorentzian for the FoG damping factor,
and we vary the maximum wavenumber included in the analysis, $k_{\rm max}$.

\begin{figure}[t]
\centering
\includegraphics[width=0.47\textwidth]{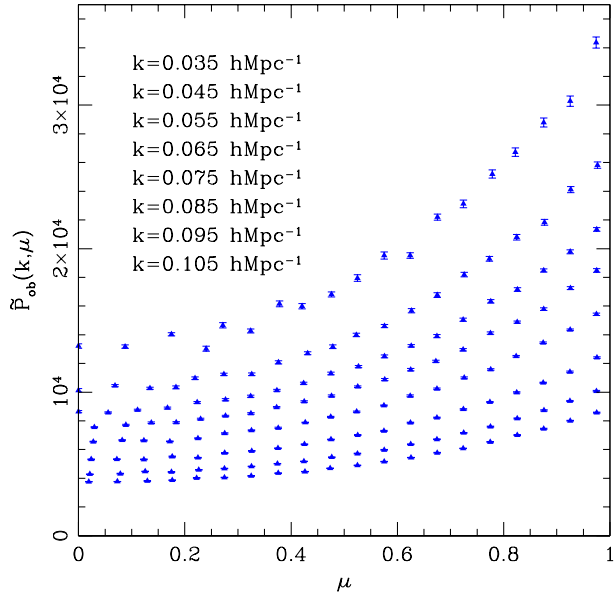}
\caption{\footnotesize The observed spectra of $\tilde{P}(k,\mu)$ are
 presented at scales $0.03\hompc < k < 0.11 \hompc$. Blue triangles
 represent averaged $\tilde{P}(k,\mu)$ at given $k$--$\mu$ bins from 60
 realizations. The error bars are dispersions of measured values of 60 realizations divided by $\sqrt{60}$.}
\label{fig:Pks}
\end{figure}

\section{Measurements of Linear Spectra}
\label{sec:measure}

\subsection{The observed spectra $\tilde{P}(k,\mu)$ from mock catalogues}

For subsequent analysis, we use the dark matter distributions created by the simulations in Ref.~\cite{Taruya:2012ut}. The volume size of the $N$-body simulations is $(2048 h^{-1}$Mpc$)^3$, and we have $60$ independent snapshots at 
each of the four output redshifts, $z=0.35$, 1, 2 and 3. The fiducial cosmological parameters of the simulation are given by $(\Omega_m=0.279, \Omega_b=0.165\,\Omega_m, \Omega_k=0, h=0.701, \sigma_8=0.816, n_s=0.96)$. The distribution of dark matter particles is modified according to their peculiar velocity to incorporate the redshift distortion effect. We adopt the distant-observer approximation and measure the power spectrum in $(k_{\bot}, k_{\|})$ space, where subscripts `$\bot$' and `$\|$' denote perpendicular and parallel components to the line-of-sight.

The density fluctuation field is constructed by assigning the dark matter particles to $1024^3$ grids for the fast Fourier transformation (FFT) using the 
cloud-in-cell (CIC)
method. We use bins in $k$ and $\mu$ for the following
analysis. $k$ is divided in $\Delta k=0.01\hompc$ linearly equally
spaced bins from $k=0.01\hompc$ to $0.2\hompc$ and $\mu$ is in 20
linear-bins from 0 to 1 with equal spacing. The averages of measured 2D
power spectra in ($k$,$\mu$) coordinate are shown in
Fig.~\ref{fig:Pks}. The Gaussian variance is used to derive errors for
each bin shown as error bars in Fig.~\ref{fig:Pks},
$\sigma[\tilde{P}_{\rm ob}(k,\mu)]=\tilde{P}_{\rm
ob}(k,\mu)\sqrt{2/N(k,\mu)}$ where $N(k,\mu)$ is number of modes in 
$2048^3 (\mpcoh)^3$ in Fourier space. 

The overall amplitude of $\tilde{P}(k_i,\mu_j)$ at $\mu\rightarrow 0$ is solely determined by $P^{\rm lin}_{\delta\delta}(k_i)$. And the running of $\tilde{P}(k_i,\mu_j)$ along $\mu$ direction is determined by $P^{\rm lin}_{\Theta\Theta}(k_i)$ about the pivot point of $\tilde{P}(k_i,\mu_j=0)$. Those distinct contributions of $P^{\rm lin}_{\delta\delta}$ and $P^{\rm lin}_{\Theta\Theta}$ to the observed spectra lead us to simultaneously decompose both through data fitting to the observed $\tilde{P}(k_i,\mu_j)$ in $k$ and $\mu$ dimension. Additionally, non--perturbative effect is externally parameterized using $\sigma_v$ appearing in $G^X$, while the higher order loop corrections are expressed using the given linear spectra parameters of $P^{\rm lin}_{\delta\delta}$ and $P^{\rm lin}_{\Theta\Theta}$. Here parameterised one-dimensional velocity dispersion $\sigma_v$ is set to be a scale-independent free parameter.

We find best-fit parameter space of $P^{\rm lin}_{gg}(k_i)$, $P^{\rm lin}_{\Theta\Theta}(k_i)$ and $\sigma_v$ by minimizing~\cite{Song:2010bk},
\ba
\chi^2&=&\sum^{i_{max}}_{i=i_{min}}\sum^{20}_{p=1} \sum^{20}_{q=1}  
[\tilde{P}_{\rm ob}(k_i,\mu_p)-\tilde{P}_{\rm fit}(k_i,\mu_p)]\nn\\
&\times&{\rm Cov}^{-1}_{pq}(k_i)
[\tilde{P}_{\rm ob}(k_i,\mu_q)-\tilde{P}_{\rm fit}(k_i,\mu_q)]\,,
\ea
where $k_{\rm min}$ is fixed to be $k_{\rm min}=0.01\hompc$, and best
$k_{\rm max}$ is determined in the following subsections. Off diagonal
elements of the covariance matrix are nearly negligible
and those diagonal elements are written as
\ba
{\rm Cov}^{-1}_{pp}(k_i)=\frac{1}{\sigma[\tilde{P}_{\rm ob}(k_i,\mu_p)]^2}\,.
\ea
We repeat this procedure for each 60 realization, and report the averages of best-fit values.

\subsection{Decomposition of linear density--density spectra}

\begin{figure}[t]
\centering
\includegraphics[width=0.47\textwidth]{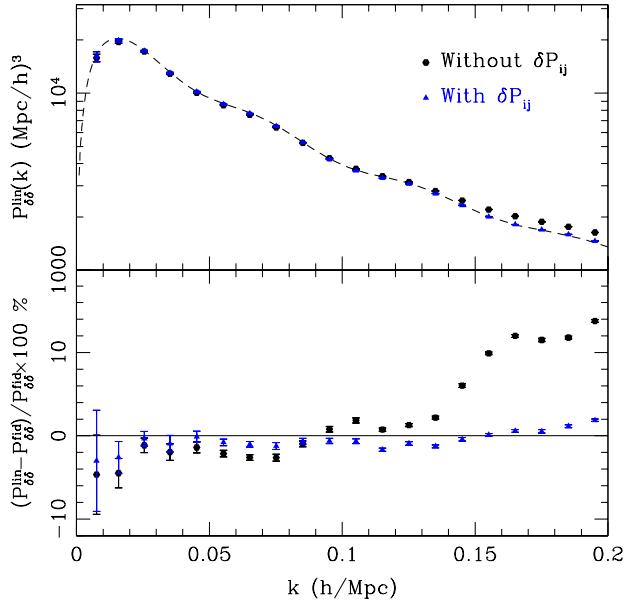}
\caption{\footnotesize Decomposed $P^{\rm lin}_{\delta\delta}$ are presented with $k_{\rm max}=0.2\hompc$. ({\it upper panel}) Dashed curve represents fiducial $P^{\rm lin}_{\delta\delta}$. Blue triangles represent fitting results using theoretical formulation in Eq.~\ref{eq:Pth}, and black circles represent fitting results without non--linear correction terms in Eq.~\ref{eq:Pth}. ({\it bottom panel}) The fractional errors are presented for measured $P^{\rm lin}_{\delta\delta}$ in the upper panel.}
\label{fig:Pkgg}
\end{figure}

We present the result of decomposition of $P^{\rm lin}_{\delta\delta}$. Linear spectra of $P^{\rm lin}_{\delta\delta}$ are determined at a few first bins about $\mu=0$ in which the orientation of correlated two point pairs is transverse. The observed $\tilde{P}_{\rm ob}$ at those $\mu$ bins are nearly equivalent to density--density spectra itself. The procedure to decompose $P^{\rm lin}_{\delta\delta}$ is immune from all line-of-sight contamination described in Sec.~\ref{sec:models}. Therefore density--density spectra are measured in high precision at arbitrary scale of $k$. But what we are interested in is measuring linear spectra of $P^{\rm lin}_{\delta\delta}$. Unless the non--linear contribution in the decomposed $P_{\delta\delta}$ are separated, cross--correlation between density and velocity fields is not guaranteed to be perfect.

The closure approximation in Eq.~(\ref{eq:closure}) is applied for extracting linear information out of measured $P_{\delta\delta}$. This approximation breaks down at specific scale of $k$. The upper bound of $k$ is investigated to uncover the limit of the closure approximation for density fields. 

We use measured $\tilde{P}_{\rm ob}(k,\mu)$ up to $k = 0.2\hompc$ at
$z=0.35$. Then there are 400 measured $\tilde{P}_{\rm ob}(k,\mu)$ at
each realization of  simulated maps. Those measured data are fitted by the
parameter space of (20$P^{\rm lin}_{\delta\delta}$, 20$P^{\rm
lin}_{\Theta\Theta}$, $\sigma_v$). In order to handle a large parameter
space, we adapt Markov chain Monte Carlo (MCMC) methods which are a
class of algorithms for sampling from probability distributions around
equilibrium points. 

Dashed curve in Fig.~\ref{fig:Pkgg} represents the fiducial $P^{\rm lin}_{\delta\delta}(k)$ in logarithmic scale. Black circle points in the top panel represent measured $P_{\delta\delta}(k_i)$ when non--linear correction terms of $\delta P_{ij}$ in Eq.~(\ref{eq:closure}) are nullified. Fractional errors against linear spectra of density fields are shown in the bottom panel of Fig.~\ref{fig:Pkgg}. Non--linearity of density fields in measured $P_{\delta\delta}(k_i)$ is observed from $k=0.1\hompc$. Measured $P^{\rm lin}_{\delta\delta}(k_i)$ deviates from $P^{\rm fid}_{\delta\delta}(k)$ by 10$\%$ at $k=0.2\hompc$. Blue triangle points represent measured $P_{\delta\delta}(k_i)$ including $\delta P_{ij}$. Seen at fractional error bars in the bottom panel, linear spectra of $P^{\rm lin}_{\delta\delta}(k)$ are well reproduced within a couple of percentage uncertainties through $k=0.2\hompc$. The closure approximation in Eq.~(\ref{eq:closure}) for density fields is proved to be trustable at least by this limit of $k$.

Cross--correlation coefficient between density and velocity fields is not parameterized in this paper. Instead, we claim to probe linear spectra of density and velocity fields using this closure approximation. Our test for linearity of measured density spectra in this subsection is important to proceed next step of probing coherent motion spectra.

Additionally, measuring $P^{\rm lin}_{\delta\delta}(k)$ is also precious information to probe distance measures. The important information about the evolution of the universe is imprinted on the large-scale structure of the universe. The broadband shape of the power spectrum provides the information about the horizon scale at the epoch of the matter--radiation equality~\cite{Gaztanaga:2008xz,Song:2012gh,Blake:2003rh,Seo:2003pu,Wang:2006qt}. Although much of this information has been faded away due to the nonlinear process, the possible extension of distance measure detectability at bigger $k$ is worth being investigated in the following works.

\begin{figure}[t]
\centering
\includegraphics[width=0.47\textwidth]{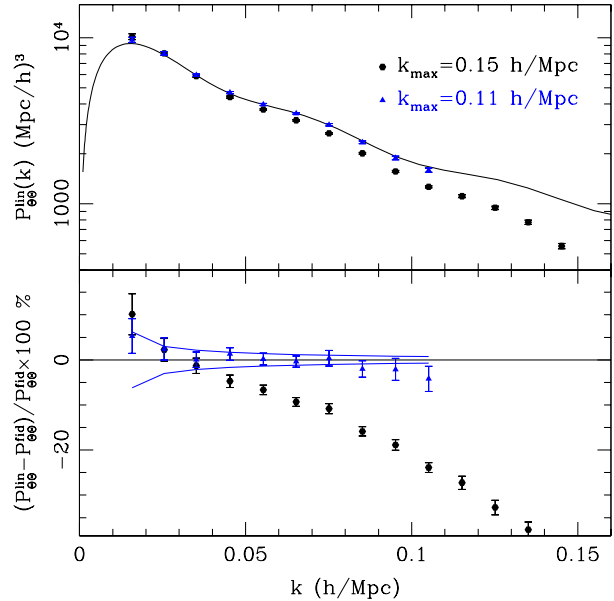}
\caption{\footnotesize Decomposed $P^{\rm lin}_{\Theta\Theta}$ are presented with varying $k_{\rm max}$. ({\it upper panel}) Solid curve represents fiducial $P^{\rm lin}_{\Theta\Theta}$. Blue triangles represent fitting results using $k_{\rm max}=0.11\hompc$, and black circles represent fitting results using $k_{\rm max}=0.15\hompc$. ({\it bottom panel}) The fractional errors are presented for measured $P^{\rm lin}_{\Theta\Theta}$ in the upper panel. Thin blue curves represent the estimated errors using the Fisher matrix analysis.}
\label{fig:Pktt}
\end{figure}

\subsection{Decomposition of coherent motion spectra}

\begin{figure}[t]
\centering
\includegraphics[width=0.49\textwidth]{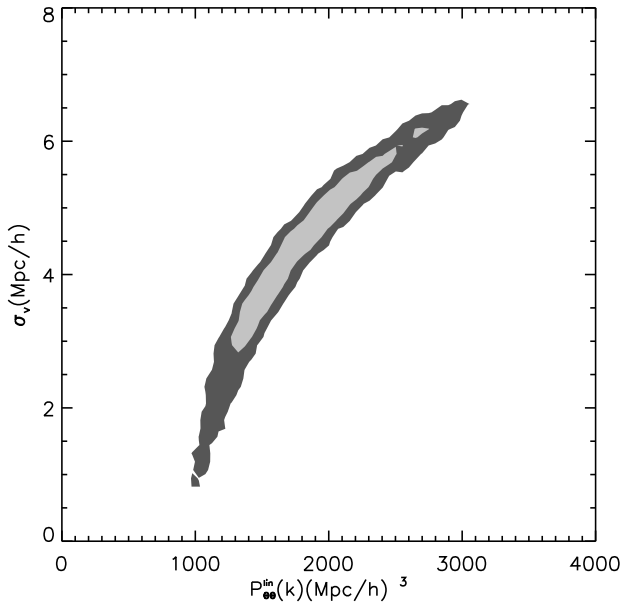}
\caption{\footnotesize Correlation between $P^{\rm lin}_{\Theta\Theta}$ and $\sigma_v$ is presented at the bin of $k=0.10\hompc$, using one realization of simulated catalogue out of 60. The upper bound of $k$ is $k_{\rm max}=0.11\hompc$. It corresponds to the last bin of blue triangle points in Fig.~\ref{fig:Pktt}.}
\label{fig:con}
\end{figure}

\begin{figure}[t]
\centering
\includegraphics[width=0.47\textwidth]{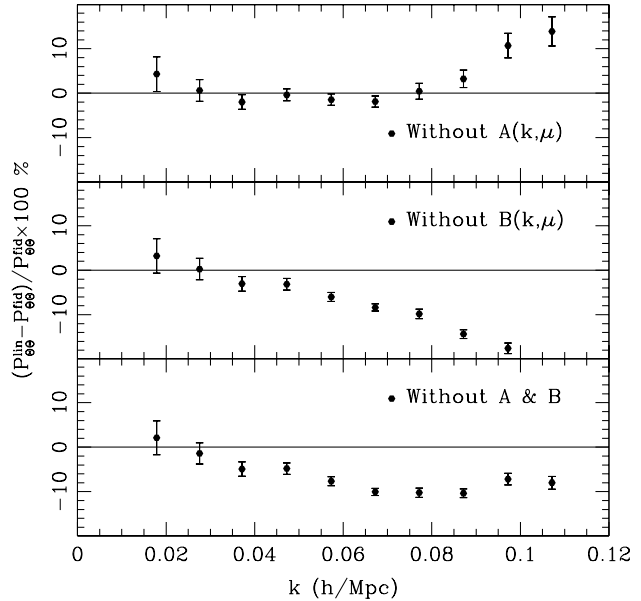}
\caption{\footnotesize We test the contribution of higher order terms of $A(k,\mu)$ and $B(k,\mu)$ in Eq.~\ref{eq:Pth}. The upper bounds of $k$ is fixed at $k_{\rm max}=0.11\hompc$. Dashed curves represent the fiducial $P^{\rm lin}_{\Theta\Theta}$, and blue triangles are fitting results using theoretical formulation in Eq.~\ref{eq:Pth}. Black circles in the upper panel represents fitting results without bi--spectral higher order terms of $A(k,\mu)$ in Eq.~\ref{eq:Pth}, black circles in the bottom panel represents fitting results without quadratic higher order terms of $B(k,\mu)$ in Eq.~\ref{eq:Pth}.}
\label{fig:PkttAB}
\end{figure}

Results of $P^{\rm lin}_{\Theta\Theta}$ decomposition are presented in this subsection. The running of $\tilde{P}_{\rm ob}$ along the $\mu$ direction is caused by $P^{\rm lin}_{\Theta\Theta}$, and pivoted from $\tilde{P}_{\rm ob}$ at $\mu=0$. The observed $\tilde{P}_{\rm ob}$ are maximally affected by peculiar velocity when the orientation of correlation is radial, while minimally affected at transverse orientation of correlation. In the limit of $\mu\rightarrow 1$, $\tilde{P}_{\rm ob}$ is significantly contaminated by all non--linear smearing effects described in~Sec.~\ref{sec:models}. Hence, we develop an appropriate statistical tool to bridge the improved theoretical model to real datasets. 

The statistical methodology to treat non--linear correction terms is described in~Sec.~\ref{sec:modelsB} in detail. In this method, the higher order loop correction terms are given theoretically, but the uncertainty due to FoG effect is phenomenologically parameterized. At scales in which the first order term of FoG dominates, the assumption of FoG effect is valid through fitting $\sigma_v$ with datasets. But beyond this quasi--linear cut--off, our statistical model is broken down. The decomposition of $P^{\rm lin}_{\Theta\Theta}$ in our analysis is limited by the uncertainty of FoG effect at higher orders.

Using the simulated maps at $z=0.35$, we find this upper bound of $k$ scales in which our assumption is valid. We test two different upper bounds of $k_{\rm max}=0.11 \hompc$ and $0.15 \hompc$. The blue triangle points in Fig.~\ref{fig:Pktt} represent the decomposed $P^{\rm lin}_{\Theta\Theta}$ using $k_{\rm max}=0.11 \hompc$. The fiducial spectra are well decomposed. The best fit $\sigma_v$ is 3.7$\mpcoh$ with $k_{\rm max}=0.11 \hompc$ which corresponds to about 10$\%$ non--perturbative contribution to $\tilde{P}_{\rm ob}$ at $\mu=1$. The detailed functional form of FoG is not crucial in our statistical method, because most FoG functions agree at the first order approximation. The decomposed $P^{\rm lin}_{\Theta\Theta}(k_i)$ is not much dependent on types of FoG function, such as Gaussian or Lorentzian, in this test. Hereafter, Gaussian function is chosen for describing FoG effect.

When we increase $k_{\rm max}$ beyond $k_{\rm max}=0.11 \hompc$, the measured $P^{\rm lin}_{\Theta\Theta}(k)$ becomes underestimated~\cite{Tang:2011qj}. The black circle points in Fig.~\ref{fig:Pktt} represents decomposed $P^{\rm lin}_{\Theta\Theta}(k)$ with $k_{\rm max}=0.15 \hompc$. This test indicates that the unknown higher order terms of FoG effect become dominating above $k=0.11 \hompc$. Beyond $k_{\rm max}=0.11 \hompc$, the first order approximation of FoG effect is not valid, and the decomposed $P^{\rm lin}_{\Theta\Theta}$ start to be biased.

We compare fractional errors of decomposed $P^{\rm lin}_{\Theta\Theta}(k)$ with the theoretical estimation using Fisher matrix analysis~\cite{White:2008jy}. We do not marginalize the Fisher matrix with FoG effect. The estimated errors using $k_{\rm max}=0.11 \hompc$ are presented as thin blue curves in Fig.~\ref{fig:Pktt}. When $P^{\rm lin}_{\Theta\Theta}(k)$ is not correlated much with FoG effect at smaller $k$ of $k<0.05\hompc$ , the observed and estimated errors agree to each other. But when $P^{\rm lin}_{\Theta\Theta}(k)$ is affected much by FoG effect at $k\rightarrow 0.1\hompc$, the observed fractional errors increase. The correlation between $P^{\rm lin}_{\Theta\Theta}(\bar{k}=0.10 \hompc)$ and $\sigma_v$ is presented in Fig~\ref{fig:con} using one realization of simulated maps out of 60. This high correlation causes the increasing fractional errors about a factor of 3.

\begin{figure}[t]
\centering
\includegraphics[width=0.47\textwidth]{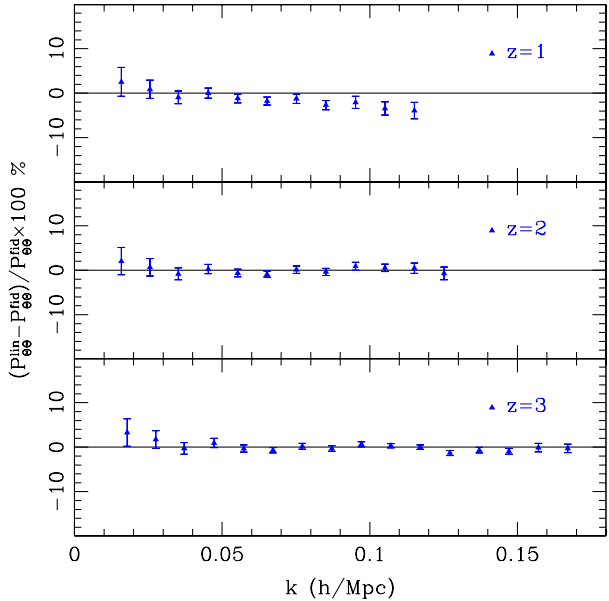}
\caption{\footnotesize Measured spectra of $P^{\rm lin}_{\Theta\Theta}(k_i)$ are presented at higher redshift at $z=1$, $z=2$ and $z=3$ in the top, middle and bottom panels respectively.}
\label{fig:Pkhighz}
\end{figure}

Results in Fig.~\ref{fig:PkttAB} present the contribution of higher--order polynomials in Eq.~(\ref{eq:Pth}). The decomposed $P^{\rm lin}_{\Theta\Theta}$ will be overestimated or underestimated without $A$ or $B$. The fractional errors in the bottom panel of Fig.~\ref{fig:PkttAB} show the results, when higher--order polynomials are not considered at all. It is the case of common practice using the simple combination of Kaiser's model and FoG effect. We show that it is not sufficient for the future precision wide--deep surveys. 

Additionally, we test our method against higher redshift data at $z=1$, $2$ and
$3$ in Fig.~\ref{fig:Pkhighz}. 
The upper bounds of $k_{\rm max}$ for these redshifts are larger than
the low redshift of $z=0.35$ as expected, since the damping scale of FoG
effect becomes rather milder at higher redshift.  We found that 
the linear velocity spectra
$P^{\rm lin}_{\Theta\Theta}$ are precisely measured up to $k_{\rm
max}=0.12$, $0.14$ and $0.18 \hompc$ at $z=1$, $2$ and $3$,
respectively. 

\section{Discussion}
\label{sec:discussion}

The decomposition results shown in previous section rely on an
idealistic assumption that the nonlinear corrections to the
redshift-space power spectra, $\delta P_{ij}$, $A$ and $B$, are known a
priori for the fiducial cosmology.  Since these corrections must be
computed for a given set of cosmological parameters, they might be 
sensitive to the underlying cosmological model and the decomposed power
spectra could be biased if we adopt the wrong cosmological priors.  In
this section, we discuss the impact of wrong cosmological assumption on
the decomposition of power spectra.

In principle, the on-going and upcoming CMB experiments will provide a 
way to precisely determine a set of cosmological parameters, from which 
we can compute the non-linear corrections. Because of 
the parameter degeneracies, however, the CMB observation alone 
cannot fully specify the cosmological model. 
This is indeed one of the main reasons why the precision measurements 
of the redshift-space distortions and/or baryon acoustic oscillations are 
highly desired to unlock the nature of late-time cosmic acceleration. 
Hence, a large statistical error of the 
cosmological parameters would cause an erroneous estimation of the 
non-linear corrections that potentially leads to a biased decomposition 
of the coherent motion spectra. To quantify the size of this, 
we consider wrong cosmological models, in which the density parameter of 
the dark energy $\Omega_v$ differs from the fiducial value 
by $3$, $5$ and $10$\%, keeping 
the equation-of-state parameter of dark energy fixed.  To be precise, 
we adopt slightly larger values of $\Omega_v$, while we fix the spatial 
curvature $\Omega_k$, the spectral index $n_s$, 
the normalization of fluctuation amplitude at CMB scales $A_s$, 
and some combinations of the parameters, $\Omega_{\rm m}h^2$ and 
$\Omega_{\rm b}h^2$, since these are expected to be 
tightly constrained by the CMB observations.  Then, we 
compute the non-linear corrections to the redshift-space power spectrum,  
and repeat the same analysis as examined in previous section.

\begin{figure}[t]
\centering
\includegraphics[width=0.47\textwidth]{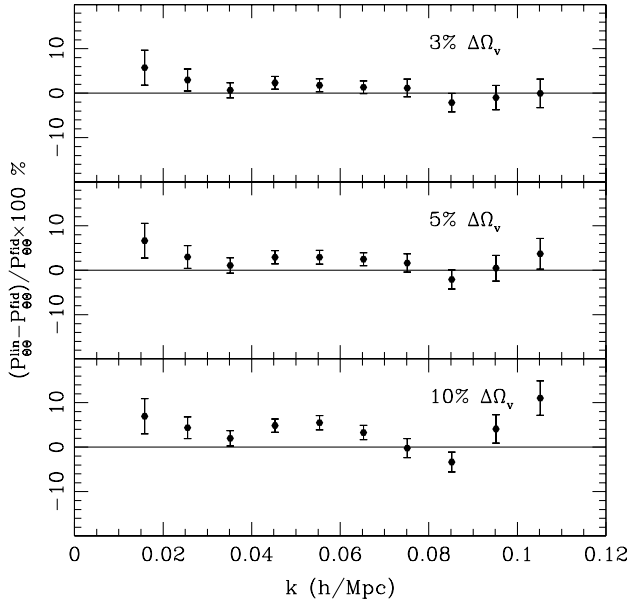}
\caption{\footnotesize Fractional errors of measured spectra of $P^{\rm lin}_{\Theta\Theta}(k_i)$ are presented at $z = 0.35$ using diverse templates at different cosmological models. Results with templates of $\Omega_v$ larger by $3\%$, $5\%$ and $10\%$ than true $\Omega_v$ are shown at top, middle and bottom panels respectively.}
\label{fig:Pktemp}
\end{figure}

Fig.~\ref{fig:Pktemp} shows the decomposition results of the coherent motion 
spectra at $z=0.35$. Here, we plot the fractional errors of the resultant 
spectra in the cases adopting  larger values of $\Omega_v$. 
As anticipated, there appears a clear systematic trend that as increasing 
$\Omega_v$, the coherent 
motion spectra tend to deviate from the fiducial spectrum. 
Nevertheless, apart from $k\sim0.1\,h$Mpc$^{-1}$, 
the resultant size of the bias is rather small. 
This is because we are 
basically looking at the scales where the contribution of the corrections 
terms is small, and even the $10$\% change of the cosmological parameter
gives a little effect on the non-linear corrections. Also, a slight 
mismatch of the non-linear corrections can be partly absorbed into the FoG 
damping factor, which further reduces the impact of wrong prior assumptions.  
Since the uncertainty of the future constraint on $\Omega_v$ will not 
be as large as $10\%$, the results shown in Fig.~\ref{fig:Pktemp} may be
regarded as a very good news, suggesting that the coherent motion spectra 
would be decomposed successfully in a less biased manner.

\section{Conclusion}
\label{sec:conclusion}

We have presented an improved prescription to 
reconstruct coherent motion spectra from the matter power spectrum in redshift 
space, properly taking account of the 
non--linear effects of both the structure growth and redshift distortions. 
Statistical analysis is presented to bridge theoretical models of redshift distortion to real dataset. Based on the perturbation theory treatment, non--linear correction terms of higher-order polynomials and non--linear growth functions are appropriately included in the reconstruction analysis. Those contributions are proved to be influential even at linear scales, and measurements of coherent motion spectra are misled without it. On the other hand, non--perturbative correction term such as FoG effect is parameterized. Those all knowns and unknowns in our analysis successfully reproduce fiducial spectra of coherent motion up to some limited scales. Although it is still steps away to achieve complete and practical observational tools measuring coherent motion spectra, results show that our analysis method righteous path to be developed to meet real world.

\acknowledgements{
Numerical calculations were performed by using a high
performance computing cluster in the Korea Astronomy and Space Science
Institute, Cray XT4 at Center for Computational Astrophysics, CfCA, of National
Astronomical Observatory of Japan, and under the Interdisciplinary Computational
Science Program in Center for Computational Sciences, University of Tsukuba.
This work is supported in part by a Grant-in-Aid for Scientific Research from the Japan Society for the Promotion of Science (JSPS) 
(No.~24740171 for I.K and No.~24540257 for A.T). 
T. N. is supported by a Grant-in-Aid for JSPS Fellows (PD: 22-181) and 
by World Premier International Research Center Initiative
(WPI Initiative), MEXT, Japan.}


\begin{thebibliography}{*}

\bibitem{Riess:1998cb} 
  A.~G.~Riess {\it et al.}  [Supernova Search Team Collaboration],
  Astron.\ J.\  {\bf 116}, 1009 (1998)
  [astro-ph/9805201].

\bibitem{Perlmutter:1998np} 
  S.~Perlmutter {\it et al.}  [Supernova Cosmology Project Collaboration],
  Astrophys.\ J.\  {\bf 517}, 565 (1999)
  [astro-ph/9812133].

\bibitem{Zhang:2007nk} 
  P.~Zhang, M.~Liguori, R.~Bean and S.~Dodelson,
  Phys.\ Rev.\ Lett.\  {\bf 99}, 141302 (2007)
  [arXiv:0704.1932 [astro-ph]].

\bibitem{Jain:2007yk} 
  B.~Jain and P.~Zhang,
  Phys.\ Rev.\ D {\bf 78}, 063503 (2008)
  [arXiv:0709.2375 [astro-ph]].
  
\bibitem{Song:2008vm} 
  Y.~-S.~Song and K.~Koyama,
  JCAP {\bf 0901}, 048 (2009)
  [arXiv:0802.3897 [astro-ph]].
 
\bibitem{Reyes:2010tr} 
  R.~Reyes, R.~Mandelbaum, U.~Seljak, T.~Baldauf, J.~E.~Gunn, L.~Lombriser and R.~E.~Smith,
  Nature {\bf 464}, 256 (2010)
  [arXiv:1003.2185 [astro-ph.CO]].
 
\bibitem{Song:2010fg} 
  Y.~-S.~Song, G.~-B.~Zhao, D.~Bacon, K.~Koyama, R.~C.~Nichol and L.~Pogosian,
  Phys.\ Rev.\ D {\bf 84}, 083523 (2011)
  [arXiv:1011.2106 [astro-ph.CO]].
 
\bibitem{Yoo:2012vm} 
  J.~Yoo and U.~Seljak,
  Phys.\ Rev.\ D {\bf 86}, 083504 (2012)
  [arXiv:1207.2471 [astro-ph.CO]].
 
\bibitem{Guzzo:2008ac} 
  L.~Guzzo, M.~Pierleoni, B.~Meneux, E.~Branchini, O.~L.~Fevre, C.~Marinoni, B.~Garilli and J.~Blaizot {\it et al.},
  Nature {\bf 451}, 541 (2008)
  [arXiv:0802.1944 [astro-ph]].
  
  
\bibitem{Song:2008qt} 
  Y.~-S.~Song and W.~J.~Percival,
  JCAP {\bf 0910}, 004 (2009)
  [arXiv:0807.0810 [astro-ph]].
  
\bibitem{Wang:2007ht} 
  Y.~Wang,
  JCAP {\bf 0805}, 021 (2008)
  [arXiv:0710.3885 [astro-ph]].
  
\bibitem{Percival:2008sh} 
  W.~J.~Percival and M.~White,
  arXiv:0808.0003 [astro-ph].
  
\bibitem{White:2008jy} 
  M.~White, Y.~-S.~Song and W.~J.~Percival,
  Mon.\ Not.\ Roy.\ Astron.\ Soc.\  {\bf 397}, 1348 (2008)
  [arXiv:0810.1518 [astro-ph]].

\bibitem{McDonald:2008sh} 
  P.~McDonald and U.~Seljak,
  JCAP {\bf 0910}, 007 (2009)
  [arXiv:0810.0323 [astro-ph]].
 
\bibitem{Jeong:2006xd} 
  D.~Jeong and E.~Komatsu,
  Astrophys.\ J.\  {\bf 651}, 619 (2006)
  [astro-ph/0604075].
  
\bibitem{Jeong:2008rj} 
  D.~Jeong and E.~Komatsu,
  Astrophys.\ J.\  {\bf 691}, 569 (2009)
  [arXiv:0805.2632 [astro-ph]].
  
\bibitem{Desjacques:2009kt} 
  V.~Desjacques and R.~K.~Sheth,
  Phys.\ Rev.\ D {\bf 81}, 023526 (2010)
  [arXiv:0909.4544 [astro-ph.CO]].
    
\bibitem{Jennings:2010uv} 
  E.~Jennings, C.~M.~Baugh and S.~Pascoli,
  Mon.\ Not.\ Roy.\ Astron.\ Soc.\  {\bf 410}, 2081 (2011)
  [arXiv:1003.4282 [astro-ph.CO]].
  
\bibitem{Reid:2011ar} 
  B.~A.~Reid and M.~White,
  arXiv:1105.4165 [astro-ph.CO].
  
\bibitem{Okumura:2011pb} 
  T.~Okumura, U.~Seljak, P.~McDonald and V.~Desjacques,
  JCAP {\bf 1202}, 010 (2012)
  [arXiv:1109.1609 [astro-ph.CO]].
  
\bibitem{Kwan:2011hr} 
  J.~Kwan, G.~F.~Lewis and E.~V.~Linder,
  Astrophys.\ J.\  {\bf 748}, 78 (2012)
  [arXiv:1105.1194 [astro-ph.CO]].

\bibitem{Samushia:2011cs} 
  L.~Samushia, W.~J.~Percival and A.~Raccanelli,
  Mon.\ Not.\ Roy.\ Astron.\ Soc.\  {\bf 420}, 2102 (2012)
  [arXiv:1102.1014 [astro-ph.CO]].

\bibitem{Blake:2011rj} 
  C.~Blake, S.~Brough, M.~Colless, C.~Contreras, W.~Couch, S.~Croom, T.~Davis and M.~J.~Drinkwater {\it et al.},
  Mon.\ Not.\ Roy.\ Astron.\ Soc.\  {\bf 415}, 2876 (2011)
  [arXiv:1104.2948 [astro-ph.CO]].
  
\bibitem{Zhang:2012yt} 
  P.~Zhang, J.~Pan and Y.~Zheng,
  arXiv:1207.2722 [astro-ph.CO].
    
   
\bibitem{Kaiser:1987qv} 
  N.~Kaiser,
  Mon.\ Not.\ Roy.\ Astron.\ Soc.\  {\bf 227}, 1 (1987).
  
\bibitem{Fisher:1994ks} 
  K.~B.~Fisher,
  Astrophys.\ J.\  {\bf 448}, 494 (1995)
  [astro-ph/9412081].
  
\bibitem{Scoccimarro:2004tg} 
  R.~Scoccimarro,
  Phys.\ Rev.\ D {\bf 70}, 083007 (2004)
  [astro-ph/0407214].
  
\bibitem{Taruya:2010mx} 
  A.~Taruya, T.~Nishimichi and S.~Saito,
  Phys.\ Rev.\ D {\bf 82}, 063522 (2010)
  [arXiv:1006.0699 [astro-ph.CO]].
 
\bibitem{Song:2010bk} 
  Y.~-S.~Song and I.~Kayo,
  Mon.\ Not.\ Roy.\ Astron.\ Soc.\  {\bf 407}, 1123 (2010)
  [arXiv:1003.2420 [astro-ph.CO]].
\bibitem{Tang:2011qj} 
  J.~Tang, I.~Kayo and M.~Takada,
 Mon.\ Not.\ Roy.\ Astron.\ Soc.\  {\bf416}, 2291 (2011)
  arXiv:1103.3614 [astro-ph.CO].
\bibitem{Taruya:2007xy} 
  A.~Taruya and T.~Hiramatsu,
  arXiv:0708.1367 [astro-ph].


\bibitem{Taruya:2012ut}
  A.~Taruya, F.~Bernardeau, T.~Nishimichi, and S. ~Codis,
  arXiv:1208.1191 [astro-ph].

\bibitem{Gaztanaga:2008xz} 
  E.~Gaztanaga, A.~Cabre and L.~Hui,
  Mon.\ Not.\ Roy.\ Astron.\ Soc.\  {\bf 399}, 1663 (2009)
  [arXiv:0807.3551 [astro-ph]].
  
\bibitem{Blake:2003rh} 
  C.~Blake and K.~Glazebrook,
  Astrophys.\ J.\  {\bf 594}, 665 (2003)
  [astro-ph/0301632].
  
\bibitem{Seo:2003pu} 
  H.~-J.~Seo and D.~J.~Eisenstein,
  Astrophys.\ J.\  {\bf 598}, 720 (2003)
  [astro-ph/0307460].
  
\bibitem{Wang:2006qt} 
  Y.~Wang,
  Astrophys.\ J.\  {\bf 647}, 1 (2006)
  [astro-ph/0601163].


\bibitem{Song:2012gh} 
  Y.~-S.~Song,
  arXiv:1210.6596 [astro-ph.CO].


\end{thebibliography}
\end{document}